\documentclass[
aps,%
11pt,%
final,%
notitlepage,%
oneside,%
onecolumn,%
nobibnotes,%
nofootinbib,%
superscriptaddress,%
noshowpacs,%
centertags]%
{revtex4}

\usepackage{multirow}

\begin{document}

\selectlanguage{english}

%\title{QUENCHED GALAXIES IN CLUSTERS OF GALAXIES AND THEIR OUTSKIRTS}
%\authorrunning{KOPYLOVA and KOPYLOV}

%\toctitle{Quenched Galaxies in Clusters of Galaxies and their
%Outskirts} \tocauthor{F.~G.~Kopylova, A.~I.~Kopylov}

\title{QUENCHED GALAXIES IN CLUSTERS OF GALAXIES AND THEIR OUTSKIRTS}

\author{\firstname{F.~G.}~\surname{Kopylova}}
%\affiliation{\saoname}
%\email{flera@sao.ru}

\author{\firstname{A.~I.}~\surname{Kopylov}}
\affiliation{\saoname}

%\received{March 27, 2020} \revised{September 18, 2020} \accepted{September 18, 2020}
\onecolumngrid
\begin{center}
{\scriptsize
Original Russian Text @ F.G.~Kopylova, A.I.~Kopylov,
published in Astrofizicheskii Byulleten, 2020,\\
Vol.75, No.4, pp.424-432}
\end{center}

\begin{abstract}
Based on the SDSS data, the properties of galaxies with
quenched star formation (QGs) within the ``splashback''-radius of
galaxy clusters $R_{\rm sp}$  and beyond it have been studied. We
used a sample of 40 groups and galaxy clusters and a sample of
field galaxies at $0.02<z<0.045$. The radii $R_{\rm sp}$ were
defined from the observed integrated distribution of the number of
galaxies as a function of the squared distance from the center of
the galaxy systems. We show that in galaxy clusters 72\% of the
QGs we have found are within $R_{\rm sp}$. About 40\% of these
galaxies are late-type ones with ${frac}DeV < 0.8$. Approximately
80\% of galaxies with quenched star formation have stellar masses
in the range of $\log M_*/M_{\odot} = [10; 11]$. We found that QGs
of late types and of early types in a less degree have maximum
angular radii $R_{90,r}$ and $R_{50,r}$ near the
``splashback''-radius of groups and clusters of galaxies. Our
results confirm the assumption that in the filaments directed
toward clusters, the quenched galaxies are more massive near
the boundaries of clusters of galaxies than at the outskirts.
\end{abstract}


\footnotetext{email: flera@sao.ru}

\maketitle

\section{INTRODUCTION}
Galaxy clusters, the largest gravitationally bound objects,
increase their mass by accreting both dark matter and luminous one
(groups of galaxies and galaxies)  along filaments from the
surrounding space (see, e.g., \cite{Colberg:Kopylova_n_en}). They
demonstrate a wide range of spatial densities of galaxies:
high-density areas in the center and low-density, almost field, on
the periphery. Exploring galaxies in different environments, one
can find out its role in quenching (reducing) star formation (SF)
in galaxies~\cite{Balogh1:Kopylova_n_en, Blanton:Kopylova_n_en,
Peng:Kopylova_n_en}. At the same time, it
is possible to study the physical mechanisms operating in
clusters, which cause the SF decrease and are responsible for the
transformation of the galaxies (see, e.g., \cite{Boselli:Kopylova_n_en}).

In the inner environment of galaxy
clusters, there are various mechanisms leading to the decrease of
the gas amount in galaxies and eventually to the decrease of star
formation.
These are tidal interactions between galaxies or
between galaxies and a potential cluster well; collisions of
galaxies at high velocities (galaxy harassment); stripping of gas
as a result of frontal pressure (ram pressure stripping) of the
intergalactic environment; thermal evaporation. According to
observational data and results of modeling the main mechanism that
leads to gas reduction in galaxies of clusters is the
ram pressure stripping (e.g., \cite{Lotz:Kopylova_n_en}).
Observational data (SDSS) show that
stripping of gas in a spiral galaxy when it merges with a cluster
is a slow process~\cite{Linden:Kopylova_n_en, Well:Kopylova_n_en}. For
2--4~Gyr~\cite{Wetzel1:Kopylova_n_en} after the galaxy hits the cluster the SF
rate practically does not change, then it quickly fades. It is
found that SF quenching occurs effectively in galaxies of stellar
masses $M_* = 10^9 - 10^{11.5}~M_{\odot}$, that is, almost in each
cluster galaxy with mass greater than
$10^{13}~M_{\odot}$~\cite{Oman:Kopylova_n_en} (SDSS data + modeling). It is
also considered that SF should decrease to zero at the first
passage of the galaxy through the center of the cluster.
Observations show that the SF rate in star-forming galaxies
continuously decreases from the periphery to the center of galaxy
clusters (e.g., \cite{Linden:Kopylova_n_en, Paccagnella:Kopylova_n_en},
and in galaxy clusters it is lower than in the field (e.g.,
\cite{Gavazzi:Kopylova_n_en, Haines1:Kopylova_n_en}).
Studies of the SF rate in
the clusters' outskirts showed that galaxies with quenched star
formation are observed even beyond the virial radii of clusters.
According to SDSS and model calculations about of 40\% of galaxies
near massive systems are escaped galaxies with quenched SF, and
their orbits can reach about $2.5 R_{200m}$~\cite{Wetzel2:Kopylova_n_en},
where
$R_{200m}$ is the radius of the sphere, inside which the system
density is 200 times the mean density of the Universe. It is also
defined that some of galaxies could lose gas within low-mass
groups  falling into the  galaxy cluster (e.g., \cite{Balogh2:Kopylova_n_en}).

In galaxy clusters (SDSS data) there is a significant number of
spirals among the galaxies with quenched SF~\cite{Well:Kopylova_n_en}.
In \cite{Hamabata:Kopylova_n_en} (SDSS data and model calculations) found the
radius of galaxy cluster ($r \sim 0.6h^{-1}$~Mpc), that
bounds the area inside which red-spiral galaxies with SF quenching
appear. In galaxy clusters of the Local universe (WINGS galaxy
sample) there are also transition types---from star-forming
galaxies to passive ones~\cite{Paccagnella:Kopylova_n_en}. They are found
within $0.6 R_{200}$ and are rare in the field. When considering
clusters as laboratories, where galaxy transformations
occur~\cite{Poggianti:Kopylova_n_en, Cebrian:Kopylova_n_en} showed that
early-type galaxies
in clusters are smaller in size than field galaxies.

To determine how the system properties of galaxies with quenched
star formation change along the normalized radius (up to
$3R/R_{200}$), in our study we used a sample of 40 galaxy
clusters~\cite{Kopylova0:Kopylova_n_en,Kopylova01:Kopylova_n_en}.
It consists of clusters of
galaxies with registered x-ray emission, which are located in the
Leo and Hercules supercluster regions. Additionally, we include
nearest systems: the clusters A1367, A1656, and 8 galaxy groups
from~\cite{Kopylova1:Kopylova_n_en}.
Selecting systems in the local
Universe ($0.02<z<0.045$) we aimed to cover the maximum range of
radial velocity dispersion---from 300~km\,s$^{-1}$ to
950~km\,s$^{-1}$. For the study we selected galaxies brighter than
$M_K = -21\fm5$. which corresponds to approximately $M_r =
-18\fm3$.

In this work, we used data of the SDSS (Sloan Digital Sky Survey Data
Release 10)~\cite{Ahn:Kopylova_n_en} and 2MASS XSC (2MASX, Two-Micron
ALL-Sky Survey Extended Source Catalog~\cite{Jarrett:Kopylova_n_en})
catalogs, and NED (NASA Extragalactic Database).

The paper is organized as follows. The second section describes
the sample of galaxies with quenched SF, the outskirts of galaxy
clusters is shown in units of radius $R_{200}$. In the third
section, we consider the characteristics of QGs: galaxy
distributions by absolute  magnitude, by the parameter that
characterizes the de Vaucouleur profile contribution into the
surface brightness profile, by the parameter equal to the axis
ratio of galaxies. The change of radii limiting 90\% and 50\% of
Petrosian flux in the $r$-band  ($R_{90,r}$, $R_{50,r}$), early-
and late-type galaxies along the normalized cluster radius
galaxies up to $3R/R_{200}$ has been studied. In Conclusion the
main results are listed. We took the following values of
cosmological parameters: $\Omega_m=0.3$, $\Omega_{\Lambda}=0.7$,
$H_0=70$~km\,s$^{-1}$~Mpc$^{-1}$.

\section{SAMPLE OF GALAXIES WITH QUENCHED STAR FORMATION}

The specific star formation rate (sSFR) in a galaxy is
defined as the integrated star formation rate divided
by the stellar mass, $sSFR = SFR/M_*$. In the distribution of galaxies by the
specific star formation rate ($\log sSFR$) the minimum is usually
found, which separates the galaxies actively forming stars (active
galaxies) from galaxies that have  quenched star formation
(quenched galaxies---QGs) and galaxies without SF (passive). Using
data on $sSFR$ and galaxy stellar mass  from the SDSS\,DR10
catalog, in our papers~\cite{Kopylova0:Kopylova_n_en,
Kopylova01:Kopylova_n_en} we selected
QGs and passive galaxies, following the condition $\log sSFR <
-10.75~[{\rm yr}^{-1}]$. Distribution of galaxies by specific star
formation rate ($\log sSFR$) has a long tail extending into the
area of galaxies without SF. In this paper, we omit galaxies of
this type ($\log sSFR <-12~[{\rm yr}^{-1}]$~\cite{Oemler:Kopylova_n_en}), and
consider only QGs, i.e., galaxies with $-12 < \log sSFR <
-10.75~[{\rm yr}^{-1}]$. Groups and clusters of galaxies in our
sample have radial velocity dispersions $\sigma$ in the range of
300--950~km\,s$^{-1}$ (which corresponds to masses $M_{200} =
(0.5-14.5)\times 10^{14}~M_{\odot}$). We divided the entire
interval into $\sigma$ bins, with subsamples of $N$ objects (shown
in Fig.~\ref{Mr}--Fig.~\ref{AB} by lines of different types and
colors):
\begin{list}{}{
\setlength\leftmargin{2mm} \setlength\topsep{2mm}
\setlength\parsep{0mm} \setlength\itemsep{2mm} }
\item 800--950~km\,s$^{-1}$ or $(9-14.5)\times10^{14}~M_{\odot}$, $N=2$; %red dash-dot-dot line;
\item 600--800~km\,s$^{-1}$ or $(4-9)\times10^{14}~M_{\odot}$, $N=6$;  %green dotted line;
\item 500--600~km\,s$^{-1}$ or $(2-4)\times10^{14}~M_{\odot}$, $N=7$; %blue short-dashed line;
\item 400--500~km\,s$^{-1}$ or $(1-2)\times10^{14}~M_{\odot}$, $N=11$; %purple dashed-dotted line;
\item 300--400~km\,s$^{-1}$ or $(0.5-1)\times10^{14}~M_{\odot}$, $N=14$. %blue long-dashed line.
\end{list}

For some of the sample objects (mostly, early-type galaxies
located within the radius of $R_{200}$) there were no data on
$sSFR$ in DR10 and we estimated $sSFR$ by the $u-r$ and $g-r$
colors of the galaxies, their absolute magnitudes and by the
${frac}DeV$ parameter based on available measurements of other
galaxies. In order to compare the results of our study we also
selected two fields, practically free from galaxy clusters. They
are located  between the Hercules and Leo superclusters. The first
one has the center coordinates ($14\fh5, 35\degr$), the radius 300
arcminutes, the redshift range $0.030< z<0.045$ ($N=219$), for the
second field, with the same redshifts and radius, the center
coordinates are ($13\fh5, 5\degr$) ($N=147$)~\cite{Kopylova0:Kopylova_n_en}.
To improve the statistics of measurements we consider both fields
together.

\begin{figure}[]
\setcaptionmargin{5mm} \onelinecaptionsfalse \captionstyle{normal}
\includegraphics[scale=0.4,bb=40 90 590 650,clip,angle=-90]{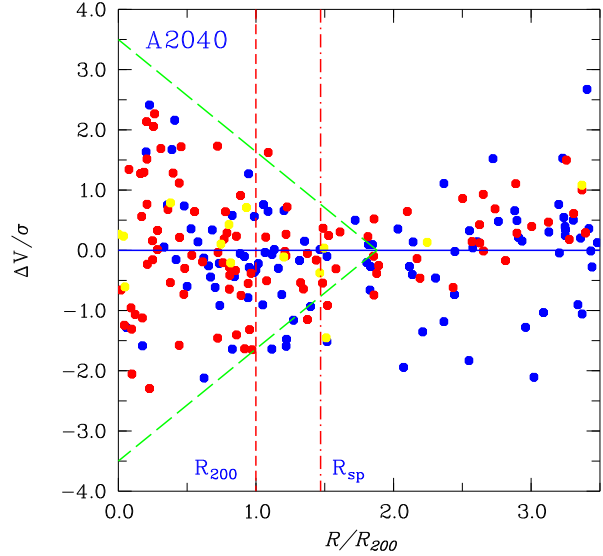}
\caption{Phase-space diagram of cluster A\,2040: velocity as a function of radius. The
velocity is ratio of the difference between the radial velocities
of galaxies and the mean radial velocity of the cluster to the
dispersion radial velocity dispersion. The $R/R_{200}$ radii are
galaxy distances from the center, normalized by the radius
$R_{200}$. Red circles show galaxies with quenched SF ($-12.0 <
\log sSFR < -10.75~[{\rm yr}^{-1}]$), yellow~--- ``passive'' ones
($\log sSFR < -12~[{\rm yr}^{-1}]$). Green model
lines~\cite{Barsanti:Kopylova_n_en} separate a cluster area with virialized
members.} \label{PPS2040}
\end{figure}

We constructed and analyzed phase-space diagrams for all objects in the
sample. One of the diagrams, for the galaxy cluster A\,2040, is
shown in Fig.~\ref{PPS2040} as an example. The radius
$R_{200}$\footnote{The radius, within which the system density is
200 times the critical density of the Universe.}, defined from the
dispersion of the radial velocities of galaxies, and the radius
$R_{\rm sp}$, which we determined from the observed cluster
profile (integrated distribution of the number of galaxies depending
on the squared distances from the cluster center~\cite{Kopylova0:Kopylova_n_en,
Kopylova01:Kopylova_n_en}) are marked. The radius $R_{\rm sp}$ (previously
denoted as $R_h$ and described in detail in~\cite{Kopylova02:Kopylova_n_en}),
is identified with the
``splashback''-radius and represents the cluster boundary, the
place of location of the orbit apocenters of the galaxies that are
flung out after the first visit to the center, i.e., the galaxies
that are already gravitationally bound to the cluster. By
estimates of~\cite{Wetzel2:Kopylova_n_en} the escaped galaxies can represent
up to 40\% of all galaxies in the region between $R_{\rm sp}$ and
$R_{200}$. Green lines (by the model of \cite{Barsanti:Kopylova_n_en}) in
Fig.~\ref{PPS2040} roughly separate the virialized part of the
galaxy cluster. Galaxies outside these lines are presumably
falling into the cluster for the first time. For other galaxy
clusters, the results of the $R_{\rm sp}$ measurements are given
in~\cite{Kopylova0:Kopylova_n_en, Kopylova01:Kopylova_n_en}. In the figures
of these papers we showed that the proportion of galaxies with quenched
star formation, QGs, decreases with increasing galaxy cluster
radii, and approaches the field values at $3R/R_{200}$.

\section{PROPERTIES OF GALAXIES WITH QUENCHED STAR FORMATION}

\subsection{Distributions of QGs by absolute magnitude, by contribution
of the de Vaucouleur profile to the surface brightness profile,
and by the axes ratio}

In 40 galaxy clusters and their surroundings we found 2368
galaxies with quenched star formation ($R/R_{200} < 3R/R_{200}$,
$M_r < -18\fm 3$). Nearly 72\%  of them are located in the inner
parts of clusters, within the normalized radius $R_{\rm sp}$. QGs
of later types, classified by parameter ${frac}DeV < $0.8,
amount to 40\%, the rest galaxies are of early-types (E, SO, Sa, and SBa).
Among the selected 2368 QGs 80\% are galaxies with stellar masses
in the range of $\log M_*/M_{\odot} = [10; 11]$, with the average
value \mbox {$\log M_*/M_{\odot} = 10.48\pm0.01$}. This
corresponds to the data of model calculations performed by
\cite{Contini:Kopylova_n_en}.

\begin{figure*}[]
\setcaptionmargin{5mm} \onelinecaptionsfalse \captionstyle{normal}
\includegraphics[scale=0.41,bb=20 0 640 550,clip]{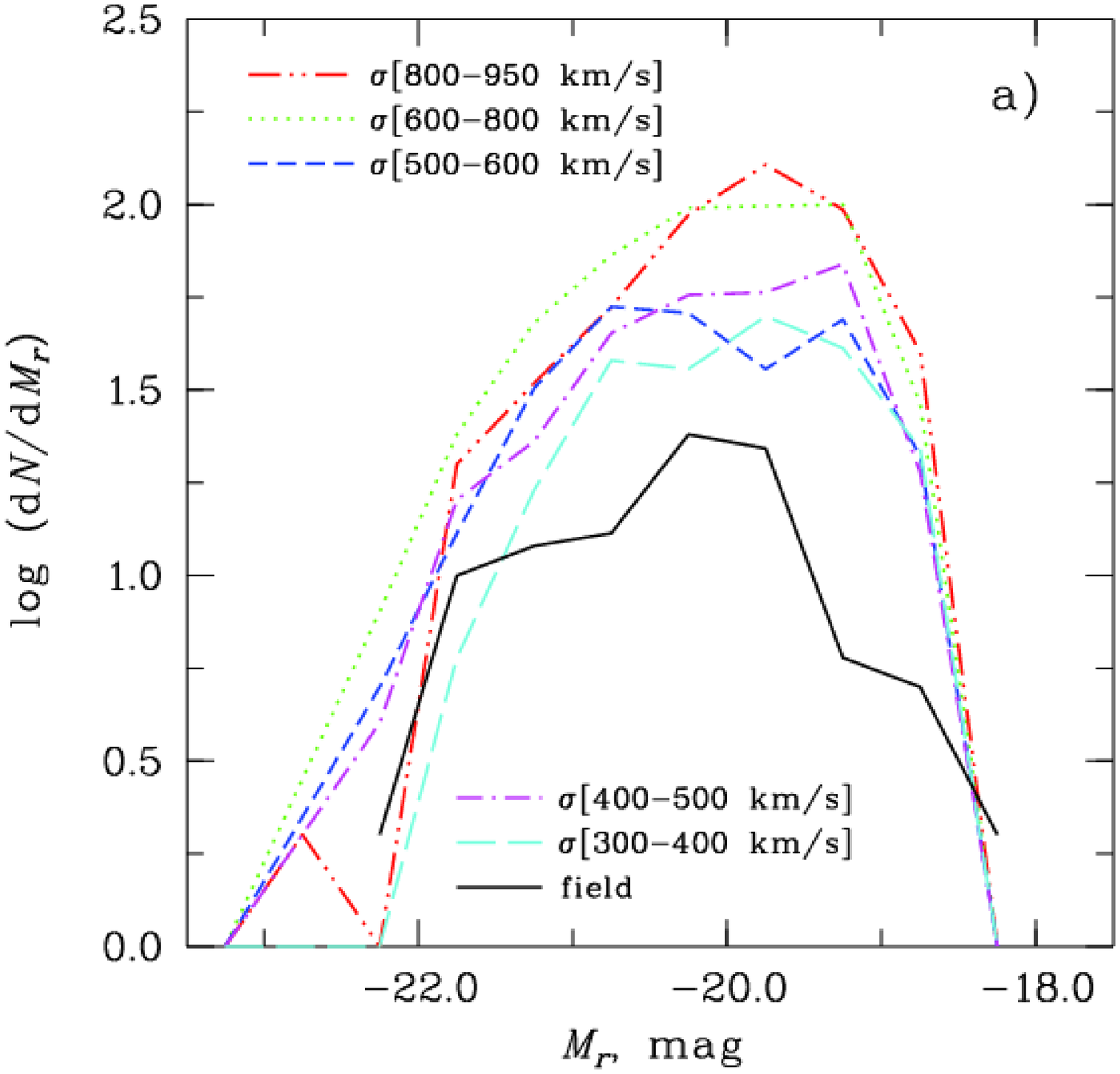}
\includegraphics[scale=0.41,bb=20 0 660 550,clip]{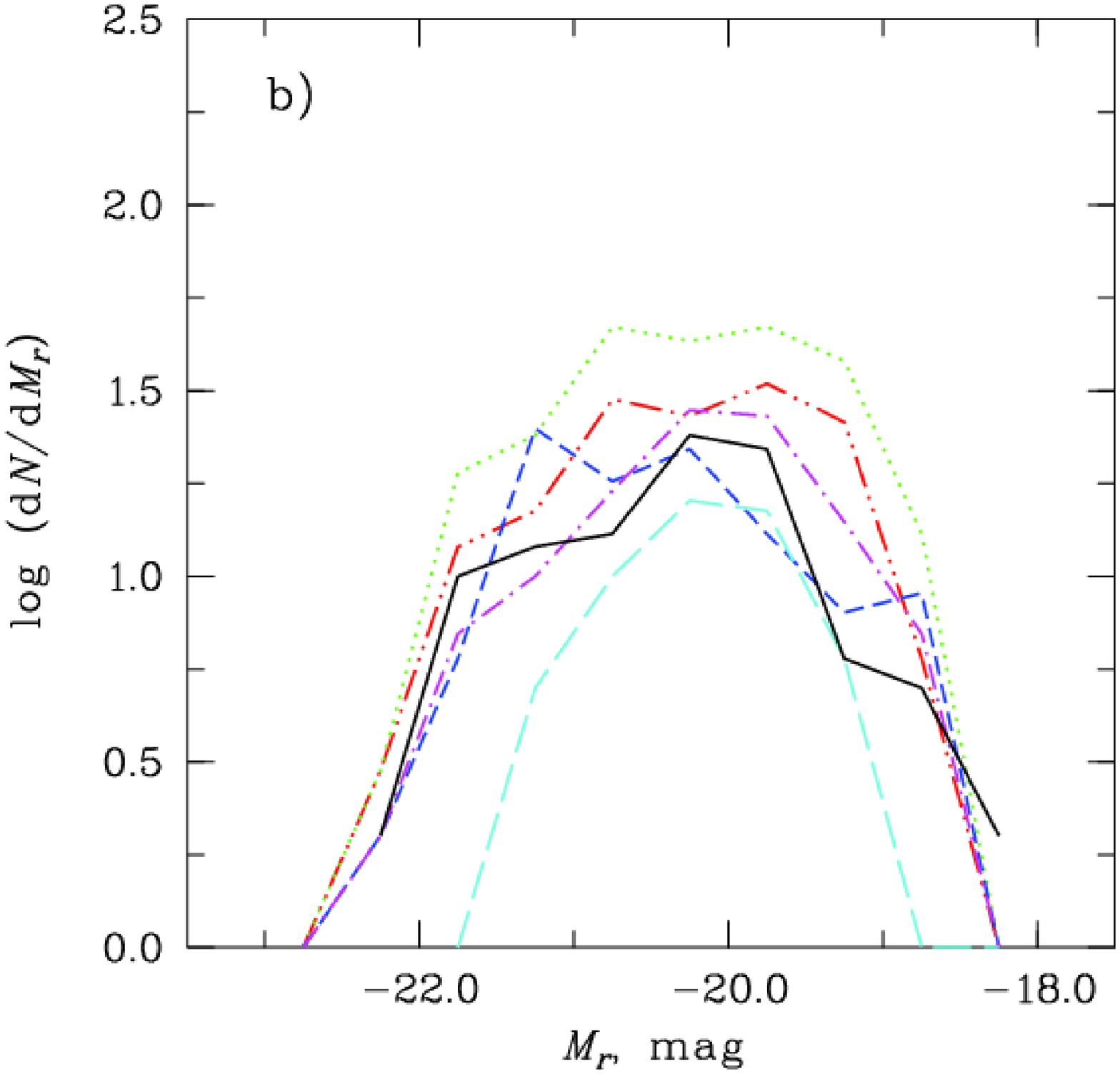}
\caption{Distributions of QGs by absolute value $M_r$ ($M_r <
-18\fm 3$): (a)~within the normalized radius $R_{\rm sp}$ and
(b)~beyond it ($R_{\rm sp}/R_{200} < R/R_{200} < 3R/R_{200}$).
Groups and clusters of galaxies are binned in accordance with the
radial velocity dispersion $\sigma$ and are shown by different
lines. The solid black line shows the distribution of galaxies in
the field.}
%The symbols are the same as in the previous drawing. %}
\label{Mr}
\end{figure*}

The distributions of QGs by absolute magnitudes within the
normalized radius $R_{\rm sp}$ and beyond it are demonstrated in
Figs.~\ref{Mr}a and~\ref{Mr}b. The errors of distributions are
small (less than 0.1, but for end points), they are estimated as
$\sqrt{dN}$, where $dN$ is the number of galaxies in the range
$dM_r$. It can be noted that in clusters ($R < R_{\rm sp}$) the
galaxies under investigation are brighter (the distribution has a
tail towards the bright ones) than in the field (black line) in
all bins except one, which corresponds to groups of galaxies with
$\sigma = 300-400$~km\,$s^{-1}$ (blue line). Outside $R_{\rm sp}$
the same groups show the absolute magnitude variations in a
smaller range than galaxies in clusters and the field ones. The
average value of the absolute magnitude of the QGs sample within
$R_{\rm sp}$ is \mbox{$\langle M_r \rangle = -20\fm 09\pm 0.02$}.
Outside of $R_{\rm sp}$ and in the field QGs are brighter:
$\langle M_r \rangle = -20\fm 27\pm 0.03$ and $\langle M_r \rangle
= -20\fm 35\pm 0.09$, respectively.

\begin{figure*}[]
\setcaptionmargin{5mm} \onelinecaptionsfalse \captionstyle{normal}
\includegraphics[scale=0.41,bb=20 0 650 550,clip]{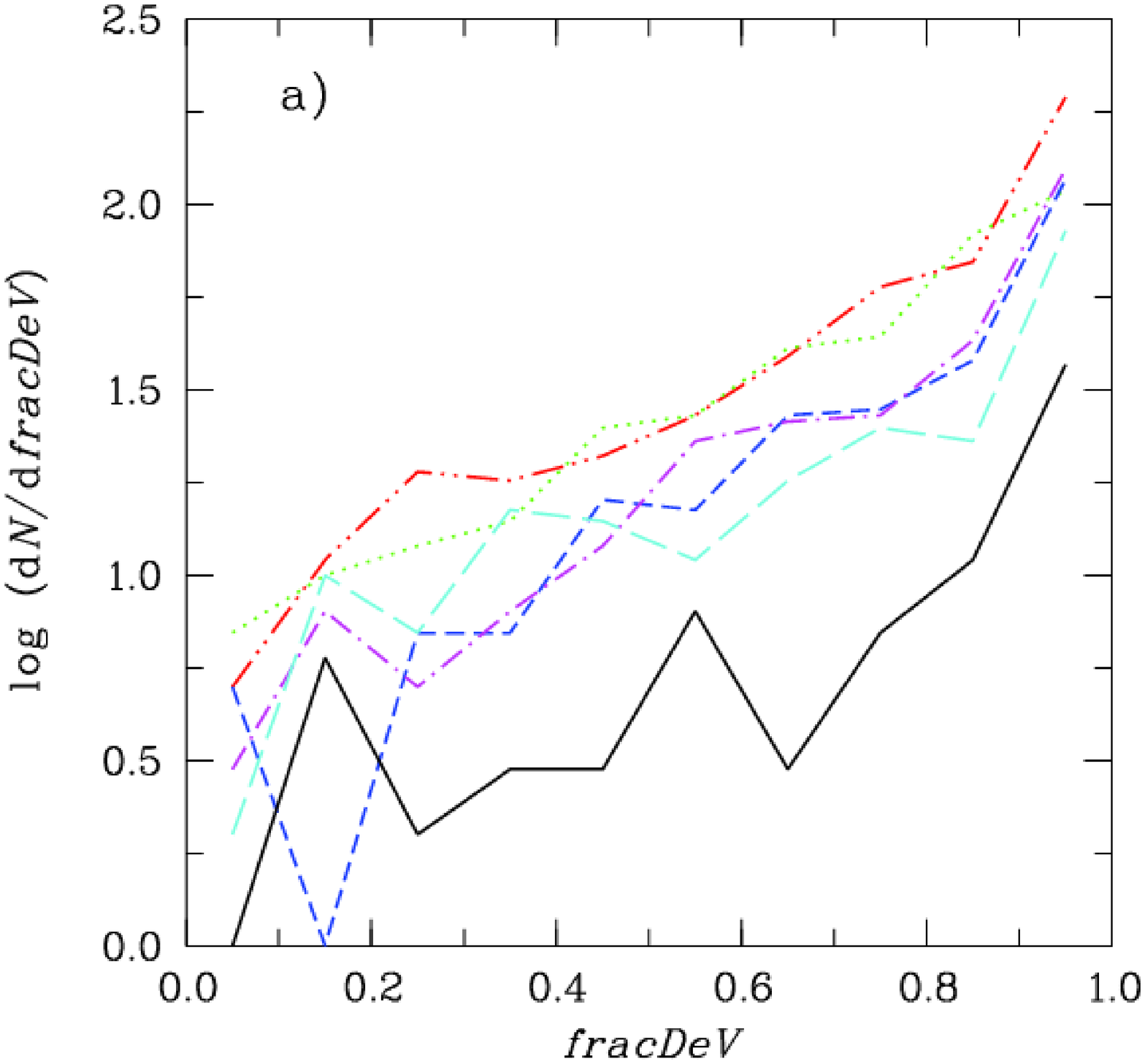}
\includegraphics[scale=0.41,bb=20 0 660 550,clip]{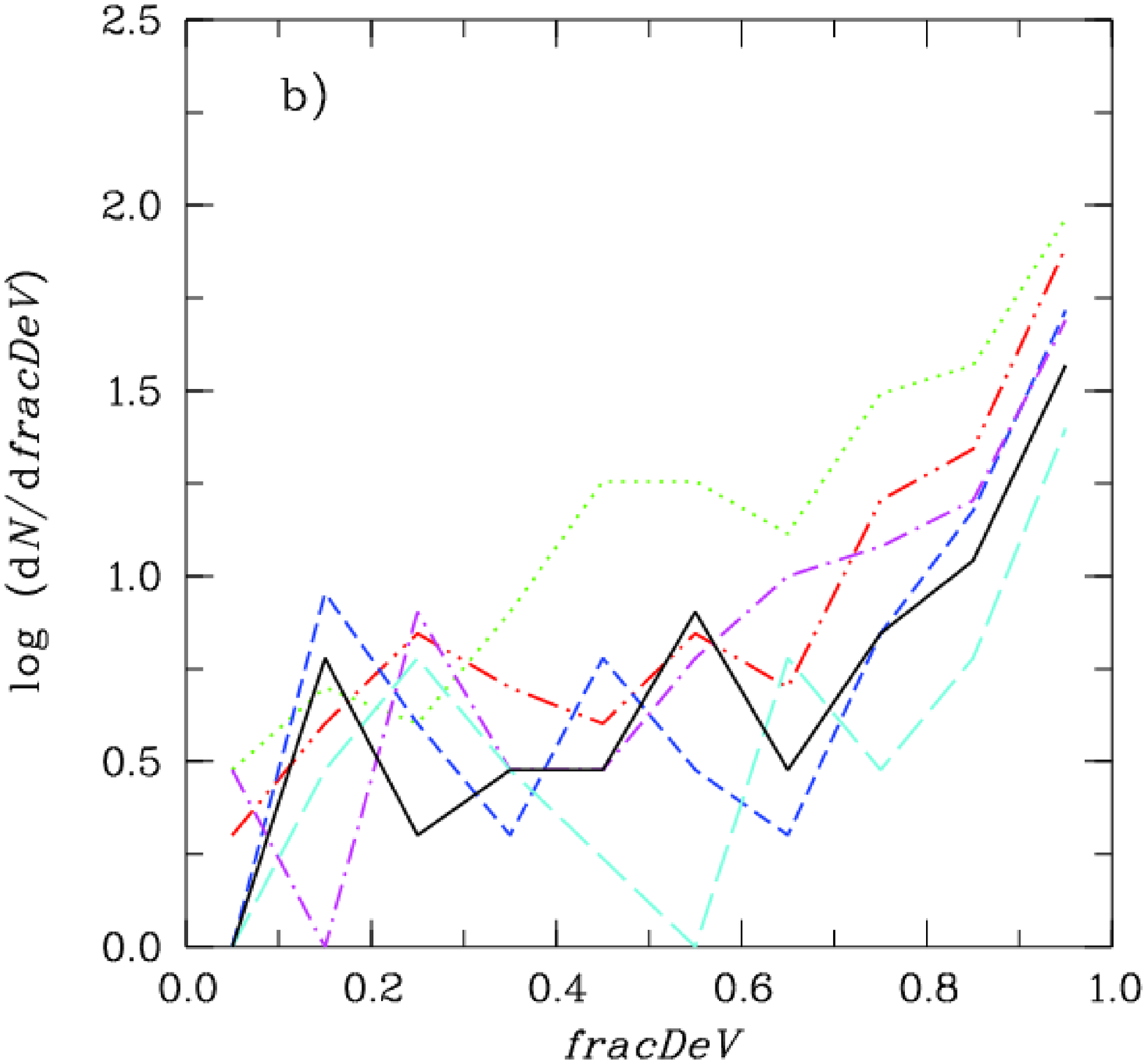}
\caption{The same as in Fig.~\ref{Mr}, but distributions by the parameter
$fracDeV$ (SDSS), describing the contribution of the de Vaucouleur
profile to the
galaxy surface brightness.} % :
%(a) within the normalized radius $R_{\rm sp}$ ($R_{\rm sp}/R_{200} < R/R_{200} <
%3R/R_{200}$) and (b) beyond it.}
\label{Vfr}
\end{figure*}

Figures~\ref{Vfr}a and~\ref{Vfr}b show distributions of galaxies
by the parameter $fracDeV$, which characterizes the contribution
of the de Vaucouleur profile to the surface brightness profile.
Note that there is no dedicated range of radial velocity
dispersions $\sigma$. In each bin there are spiral galaxies with
$fracDeV < 0.4$,  both within $R_{\rm sp}$ and beyond it, as well
as in the field. Number of QGs of early types with $fracDeV \geq
0.8$ within $R_{\rm sp}$ and beyond it are by 25\% and 38\% higher
than the number of spiral galaxies. The number of galaxies of
these types in the field is approximately the same.

\begin{figure*}[]
\setcaptionmargin{5mm} \onelinecaptionsfalse \captionstyle{normal}
\includegraphics[scale=0.41,bb=20 0 650 550,clip]{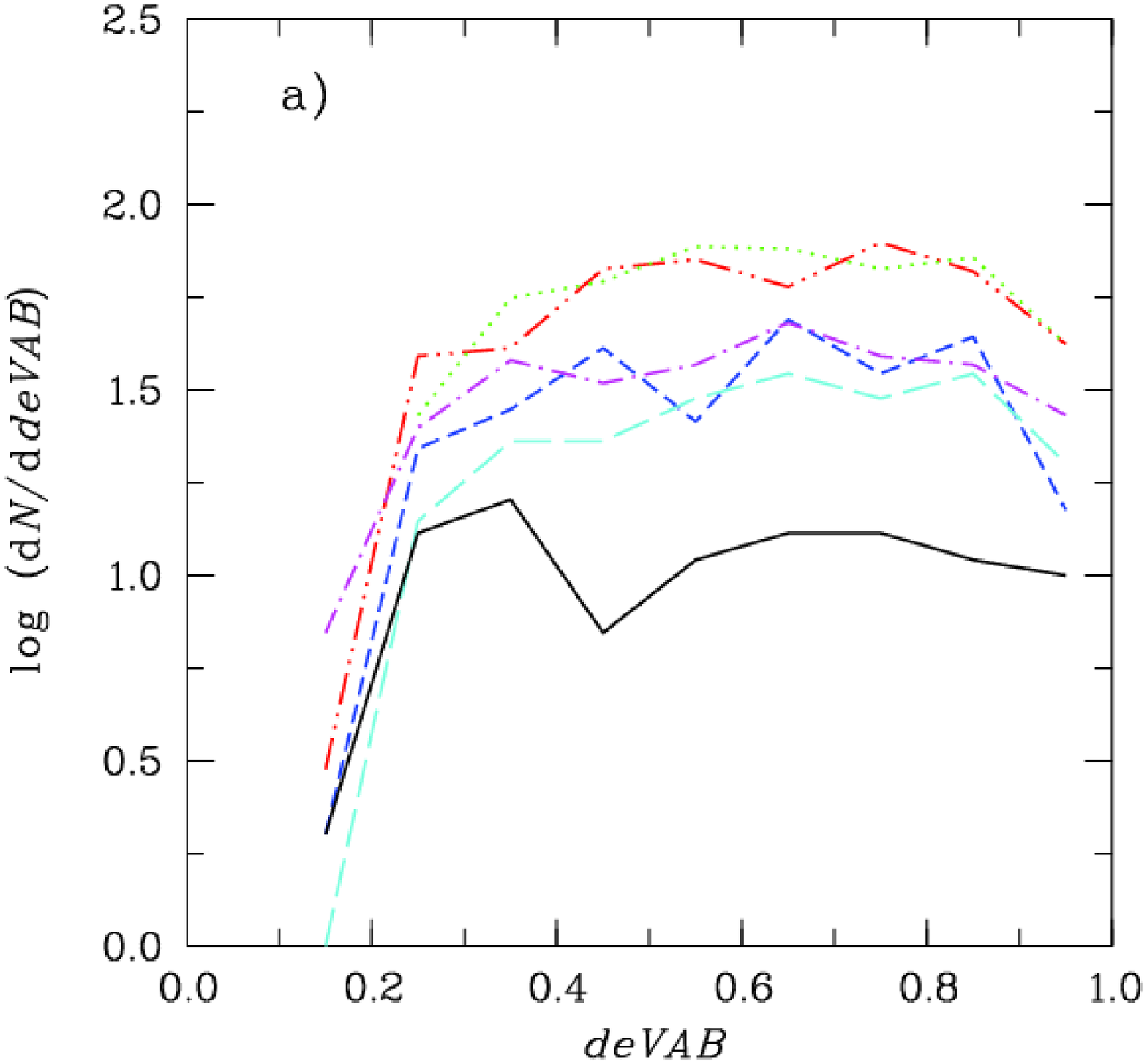}
\includegraphics[scale=0.41,bb=20 0 650 550,clip]{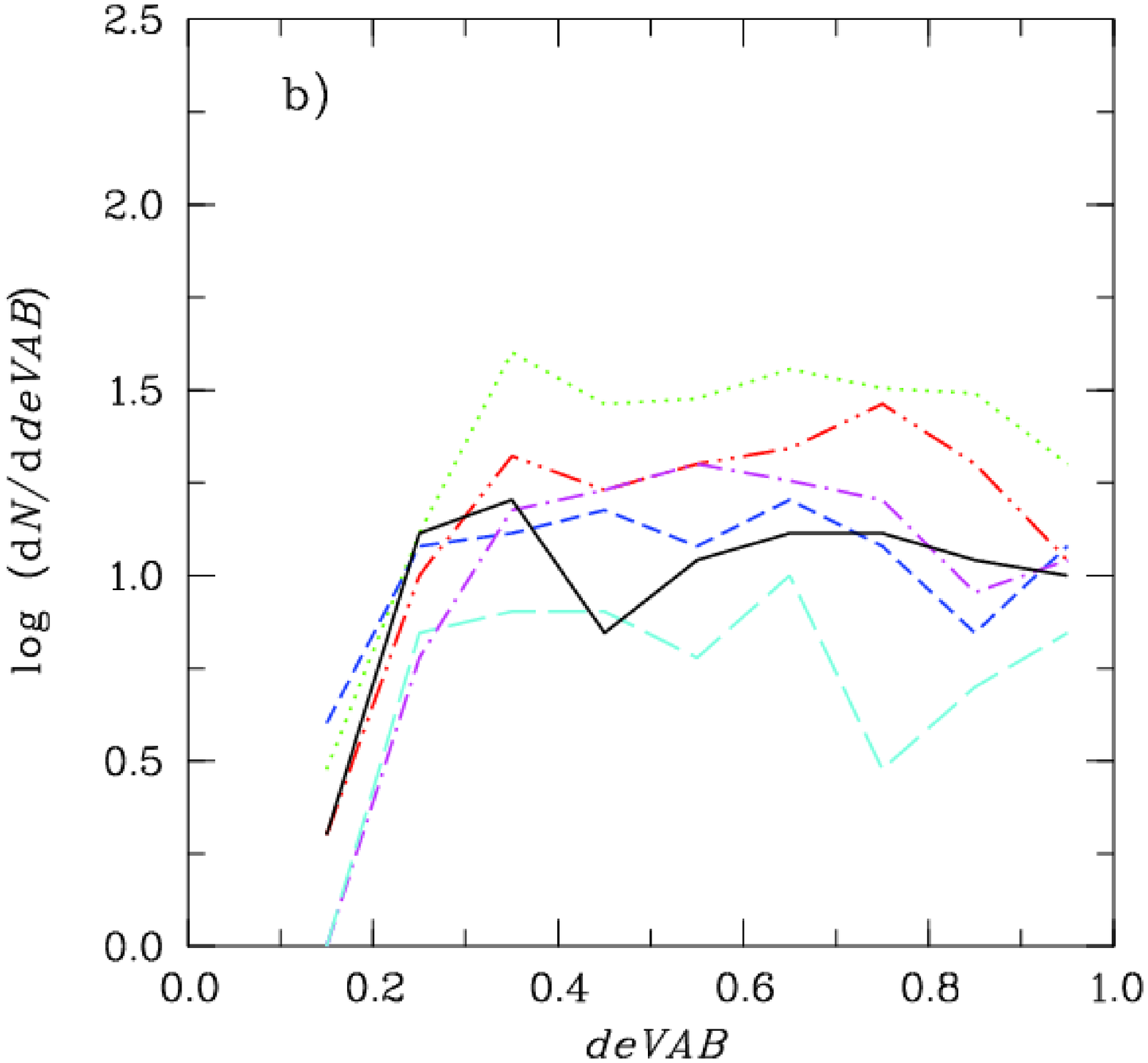}
\caption{The same as in Fig.~\ref{Mr}, but distributions
by the parameter $DeVAB$, that characterizes
the galaxy axes ratio. %: (a) within the normalized radius $R_{\rm sp}$
%($R_{\rm sp}/R_{200} < R/R_{200} < 3R/R_{200}$) and (b) beyond it.
}
\label{AB}
\end{figure*}

The distribution of the axes ratio $b/a$ for the observed surface
brightness profiles of QGs within and beyond the radius of $R_{\rm
sp}$ is  shown in Figs.~\ref{AB}a and~\ref{AB}b. It can be noted
that QGs have a variety of axes ratios---from 0.25 to 1.00. On
average the ratio $b/a$ is equal to $0.6\pm0.01$ for all studied
regions, and the number of galaxies with $b/a$ values greater or
less than this threshold is approximately the same in the field
and in the vicinity of galaxy clusters. But in galaxy clusters,
within $R_{\rm sp}$, the galaxies with $b/a > 0.6$ represent more
than 14\%. Besides, in Fig.~\ref{AB} one can see that spiral
galaxies with $b/a<0.4$ are particularly distinguished.

By all parameters we can make an overall conclusion that in the
vicinity of the lowest-mass galaxy groups those with quenched SF
are less in number, even in comparison with the field (in
Figs.~\ref{Mr}, \ref{Vfr}b, \ref{AB}b the blue lines are located
below the others).

\subsection{The change of QGs sizes along the radius of galaxy clusters}

In order to study how sizes of galaxies change, the comparison of
Petrosian radii can be used as a method that is independent of
galaxy distances. The SDSS catalog provides the data on $R_{50}$
and $R_{90}$, the angular Petrosian radii in all SDSS-bands,
containing 50\% and 90\% of a Petrosian flux. The Petrosian flux
in any band is measured as the flux within $2R_P$, where $R_P$ is
the angular radius inside which the ratio of the local surface
brightness at a radius $r$ from the center of an object to the
mean surface brightness within $r$ is equal to 0.2. We used
measurements in the $r$-filter. According to~\cite{Andreon:Kopylova_n_en},
the
quenching rate of star formation in clusters with masses of $\log
M_{200}/M_{\odot} > 14$  (as in our sample) does not depend on the
mass, richness, and iron abundance in the cluster. In our
paper~\cite{Kopylova01:Kopylova_n_en} (Fig.~7) we have shown that the
change in
the fractions of galaxies with quenched SF at a fixed stellar mass
is the same in all mass ranges considered, except for the bin for
galaxy clusters with $\log M_*/M_{\odot} = [11; 11.5]$ and $\sigma
= 300-400$~km\,s$^{-1}$. In such systems there are practically no
massive galaxies with quenched SF outside of the virial radius.
Thus, to investigate the size distribution of galaxies along the
radius, normalized by $R_{200}$, we stacked data for all the
systems.

\begin{figure*}[]
\setcaptionmargin{5mm}
\onelinecaptionsfalse
\captionstyle{normal}
\includegraphics[scale=0.60,bb=35 50 410 740,clip,angle=-90]{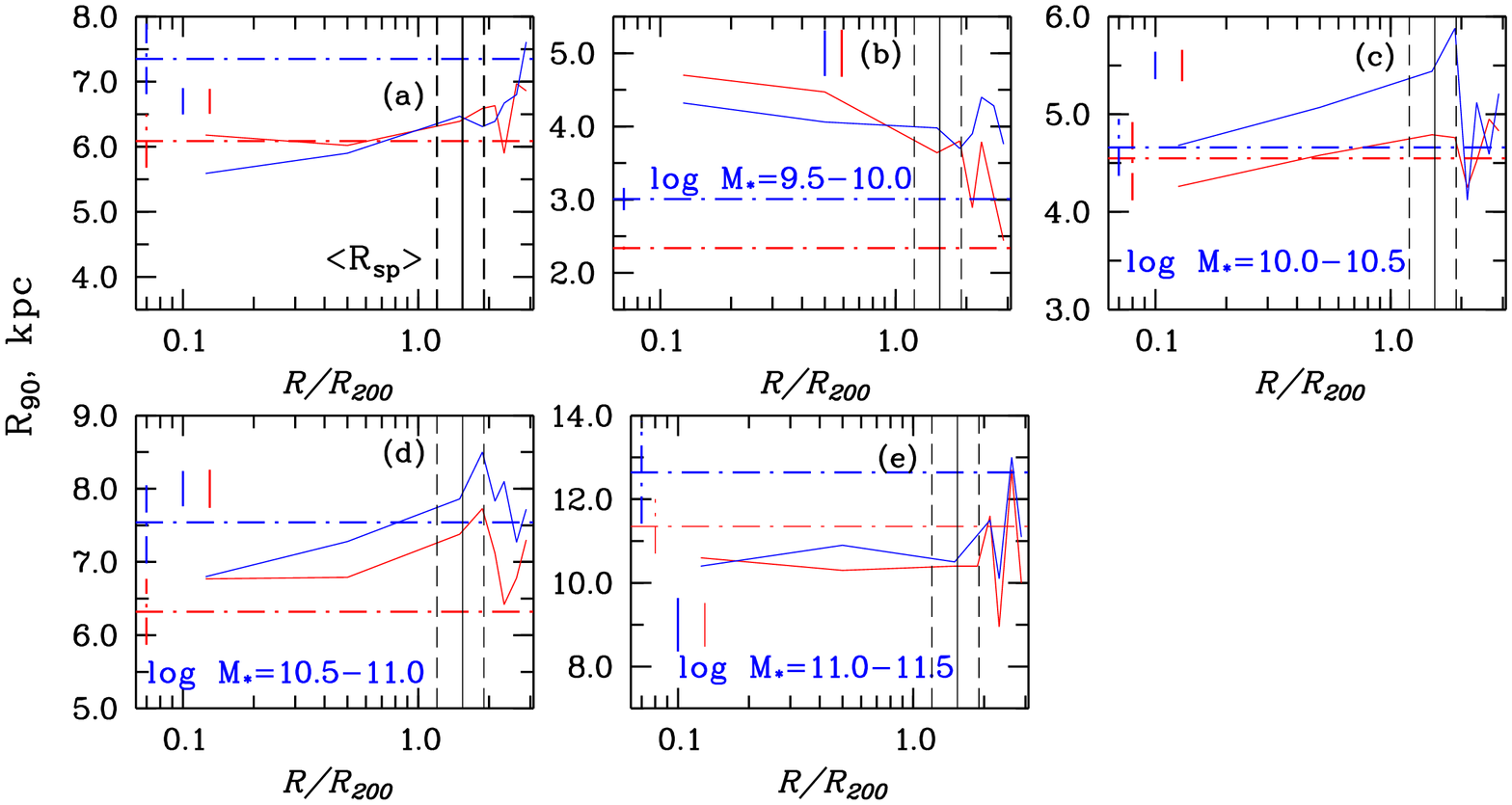}
\caption{Petrosian radius $R_{90}$ in kpc in the filter $r$
depending on the normalized radius $R/R_{200}$ for the studied
galaxies in total~(a) and binned by stellar masses~(b)--(e). Data
for galaxy clusters are stacked taking account of their normalized
radii. QGs-galaxies of early types, defined by the parameter
$fracDeV \geq 0.8$ and the late-type ones with $fracDeV < 0.8$ are
shown with red and blue solid lines, respectively. Short red and
blue lines correspond to the average errors of the radius
measurements. Solid and dashed vertical lines show the mean
$R_{\rm sp}$ and the range of its changes. Dashed horizontal lines
correspond to the average values of $R_{90}$ for the field, the
vertical ones---to the error bars.} \label{R90K}
\end{figure*}

We divided the QGs sample into two subsamples: the early-type
galaxies with $fracDeV > 0.8$ and the later-type ones with
$fracDeV \leq 0.8$. The panels in Fig.~\ref{R90K} show changes in the
average Petrosian radius $R_{90,r}$ along the normalized radius of
galaxy clusters for both of them. Similar values of the Petrosian
radius for the field galaxies are given for comparison, along with
the average radius of all galaxies in the sample $\langle R_{sp}
\rangle = (1.54 \pm 0.06)~R_{200}$ (a solid vertical line) and the
ranges of its variation $R_{sp} = [1.9; 1.2]$~\cite{Kopylova0:Kopylova_n_en,
Kopylova01:Kopylova_n_en}.

There are 58\% of early-type galaxies in the sample. The galaxies
with stellar masses in the range $\log M_* = [10; 11]$ represent
83\% among them and 74\% in the subsample of late-type galaxies.
Figures~\ref{R90K}c and~\ref{R90K}d show changes in the angular
radii $R_{90,r}$ for these galaxies. Radii of late-type galaxies
with stellar masses  in the ranges $\log M_* = [10; 10.5]$ and
$\log M_* = [10.5; 11.0]$ near $R_{\rm sp}$ are about 20\% and
13\% larger than in the central regions. For early-type galaxies
in the same stellar mass ranges the radii change is about 11\%. We
found that such an increase in angular radii for galaxies near
$R_{\rm sp}$ is also observed for $R_{50,r}$, but to a less
degree, by 13\% and 11\%, respectively. Beyond $R_{\rm sp}$ for
galaxy clusters, the angular radii of galaxies $R_{90, r}$
decrease to the values that are close to the field ones. In the
area between $R_{200}$ and $R_{\rm sp}$ different galaxies are
observed (Fig.~\ref{PPS2040}). Some of them flew out of the
cluster, and the others, moving in the opposite direction, just
come up to it. Since the galaxy sizes in this boundary area are
larger than in the central regions of clusters, then we can assume
that here, basically, there are galaxies that are not ejected from
the system, but falling on it.

The low-mass late-type galaxies in the mass range $\log M_* =
[9.5; 10.0]$ (presented in blue in Fig.~\ref{R90K}b) show no
radius changes as well, while the early-type galaxies~(red), on
the contrary, grow in size by approximately 30\% to the center of
galaxy clusters compared to the periphery. Their sizes are bigger
than those of the field galaxies, although the data accuracy for
the field estimates is rather poor---there are few galaxies in the
sample. The most massive galaxies in the range $\log M_* = [11.0;
11.5]$ (Fig.~\ref{R90K}e) also do not show significant changes in
size, although they are still smaller than field galaxies,
especially the late-type ones. In Fig.~\ref{R90K}a, where all
studied galaxies are presented without fixing the stellar mass,
changes in the radius $R_{90,r}$ of the late-type galaxies are
shown---the galaxy radii increase by 25\% to $3R/R_{200}$ and
become of the same value as in the field. At the same time, the
early-type galaxies, which are located in the inner parts of
clusters, practically do not differ from the field galaxies, but
at $3R/R_{200}$  we observe an increase in their size by 12\%.

According to, for example,~\cite{Rhee:Kopylova_n_en}, in the inner environment
of galaxy clusters the low-mass spiral galaxies on $z=0.8$ with
$\log M_*/M_{\odot} = [9; 10]$ can better resist the direct gas
stripping process. Perhaps, these galaxies have more central gas.
As a result, their star formation is not completely quenched and
still continues for a long time after they entered the cluster,
leading to their weight and size growth.

As for massive galaxies, they have a smaller gas fraction and lose it
faster and earlier, before the cluster entering~\cite{Rhee:Kopylova_n_en}
within the filaments. Other studies of a sample of 700 galaxies ($z<0.063$,
SDSS data) show that the size of spiral galaxies in clusters is 15\%
smaller, than in the field. \cite{Poggianti:Kopylova_n_en} found that the
early-type
galaxies in clusters are also smaller in size compared to the field ones.

\section{SUMMARY AND CONCLUSIONS}

Quenching of star formation in galaxy clusters and their outskirts
is a consequence of the gas loss in haloes, disks, and inner areas
of galaxies. As a result, the galaxy does not have enough ``fuel''
to form stars. In this case, an important role is played by processes
affecting galaxies within clusters and
in groups of smaller masses in filaments stretching to galaxy
clusters. Several papers have been devoted to studying the
properties of galaxies in filaments. Using the GAMA survey,
in \cite{Alpaslan:Kopylova_n_en} obtained that isolated spiral galaxies have
higher stellar masses and lower values of the SF rate in the
central regions of filaments than at their periphery. The SDSS
data allowed \cite{Chen:Kopylova_n_en} to conclude that galaxies in filaments
are more massive than outside of them despite the local density of
the same value. Study of SF quenching in the vicinity of 14
clusters the WINGS galaxy survey showed that this process is the
most effective in the direction of filaments~\cite{Salerno:Kopylova_n_en}.
In~\cite{Santiago:Kopylova_n_en} it is also shown that in the filaments the
galaxies are more massive and the SF rate in them is lower than in
the surrounding field. It is concluded, that it's likely that the
galaxies here are experienced merging, i.e. filaments
significantly influence galaxy evolution. According to our studies
of late-type QGs,  near $R_{\rm sp}$ the galaxies with the maximum
angular radius of $R_{90,r}$ probably have already lost their gas
being a part of low-mass groups falling onto a cluster along
filaments.

The star formation rate in galaxies decreases with local
density increasing In galaxy clusters this corresponds to a
decrease in the distance from the center~\cite{Balogh1:Kopylova_n_en}. There
are several models of SF quenching in galaxy clusters. In  the
``rapid quenching'' model the SF is quenched for a short time
(less than 1~Gyr) after the galaxy has entered the cluster~\cite{
Balogh2:Kopylova_n_en}. In \cite{Linden:Kopylova_n_en,
Taranu:Kopylova_n_en}
considered models of slow SF quenching. The model, best describing
SF quenching in galaxies, is the ``delayed-then-rapid'' one. In
the framework of this model the galaxy after entering a cluster
has its SF unchanged for several gigayears, and then SF quickly
quenched~\cite{Wetzel:Kopylova_n_en,Fossati:Kopylova_n_en,Foltz:Kopylova_n_en}.

In a sample of 40 galaxy clusters ($0.02 < z < 0.045$) located
mainly in the Leo, Hercules, and Coma superclusters we selected
galaxies with quenched star formation ($-12.0 < \log sSFR <
-10.75$~[yr$^{-1}]$). To study how properties of QGs change with
their distance from the cluster center, we determined the observed
``splashback''-radius. It  shows the cluster halo boundary, i.e.
area of location of both the galaxies, that flung out after the
first system passage, and those approaching the system for the
first time along with filaments. The main part of QGs in the
sample (of about 80\%) is within the radius $R_{\rm sp}$, their
absolute magnitudes being $\langle M_r \rangle = -20\fm09
\pm0.02$. If QGs are binned by stellar mass, then 80\% of galaxies
have stellar masses in the interval $\log M_*/M_{\odot} = [10;
11]$ with the mean value $\log M_*/M_{\odot} = 10.48\pm 0.01$. The
QGs sample has also the following mean values of parameters
according to SDSS data: $\langle fracDeV \rangle = 0.76 \pm0.01$
and $\langle DeVAB \rangle = 0.61 \pm0.01$. Among the galaxies
with quenched SF, about 40\% are the late-type ones located within
$R_{\rm sp}$. There are 63\% of late-type galaxies with an axis
ratio of $b/a < 0.6$ within $R_{\rm sp}$ and 72\%---beyond it.

Radii of late-type galaxies in the stellar mass range $\log M_* =
[10; 10.5]$ and $\log M_* = [10.5; 11.0]$ are by approximately
20\% and 13\% larger near the radius $R_{\rm sp}$ than in the
central areas. Radii of early-type galaxies for the same stellar
mass ranges have changed by about 11\%. Beyond
``splashback''-radius of galaxy clusters $R_{\rm sp}$  the angular
radii $R_{90,r}$ of galaxies decrease to values close to the field
ones. The main results of our work can be formulated as follows.
Sizes of the main amount of QGs galaxies of late types (and
early-type galaxies, to a lesser degree), their Petrosian angular
radii $R_{90,r}$, $R_{50,r}$ are maximum near the boundary of galaxy
clusters, defined by the ``splashback''-radius  $R_{\rm sp}$.

\begin{acknowledgments}
This research has made use of the NASA/IPAC Extragalactic Database
(NED, \url{http://nedwww.ipac.caltech.edu}), which is operated by
the Jet Propulsion Laboratory, California Institute of Technology,
under contract with the National Aeronautics and Space
Administration, Sloan Digital Sky Survey (SDSS,
\url{http://www.sdss.org}), which is supported by Alfred P. Sloan
Foundation, the participant institutes of the SDSS collaboration,
National Science Foundation, and the United States Department of
Energy and Two Micron All Sky Survey (2MASS,
\url{http://www.ipac.caltech.edu/2mass/releases/allsky/})
\end{acknowledgments}

\section*{CONFLICT OF INTEREST}
The authors declare no conflict of interest regarding this paper.

\begin{center}
\refname
\end{center}

\end{document}